\documentstyle[11pt,paspconf,times]{article}

\def\cwcpltmacro#1#2#3#4#5#6#7{\centering \leavevmode
    \vbox to#2{\rule{0pt}{#2}}
    \includegraphics{#1}}

\begin{document}

\title{Kinematics and Cloud to Cloud Abundance Ratios: A Rosetta Stone
for Disk and Halo Components in Mg II Absorbers}

\author{Christopher~W.~Churchill and Jane~C.~Charlton}
\affil{Pennsylvania State University, University Park,PA 16802}

\begin{abstract}
Based upon QSO absorption line studies, galaxies at $0.3 \leq z \leq
1.0$ are seen to have extended gaseous ``halos''.  When the absorption
lines are observed at high resolution, the velocity distribution of
the absorbing gas gives clues as to where this gas actually arises in
and around the galaxies.  Cloud to cloud abundance ratios provide
further clues, given that the ionization conditions and chemical
evolution histories of outer halos clouds, spiral disk clouds, and
clouds in elliptical galaxies are different.  The key to
interpreting the high redshift data is the Milky Way galaxy, the 
absorption line ``Rosetta stone''.
\end{abstract}

\section{Introduction}

In this volume, Steidel has described observational evidence for
extended gaseous ``halos'' surrounding normal $L^{\ast}$ galaxies
at $0.3 \leq z \leq 1.0$ (also see Steidel 1995), where the tracer for
low ionization conditions is the strong resonant {\rm Mg}\kern
0.1em{\sc ii}~$\lambda\lambda 2976, 2803$ doublet.
Mo has reviewed the properties of these halos from a theoretical view
point (this volume).
Here, we discuss methods for constraining the spatial distribution of
the absorbing gas using high resolution data and models based upon
known spatial distributions, kinematics, and chemical enrichment
histories of galactic gas.
Is there a relationship between the gas kinematics and the spatial
locations/distributions of the absorbing gas that can be exploited to
infer where in a galaxy absorbing gas arises? 
Is there a relationship between metal abundance patterns and cloud
velocities that could be exploited to this end?
It may be that, on a case--by--case basis, we can infer what parts of
galaxies are giving rise to the absorbing gas.
With a large enough sample we could then infer details about the
evolution of galactic gas.

\section{Observations and Models}

\begin{figure}[th]
\cwcpltmacro{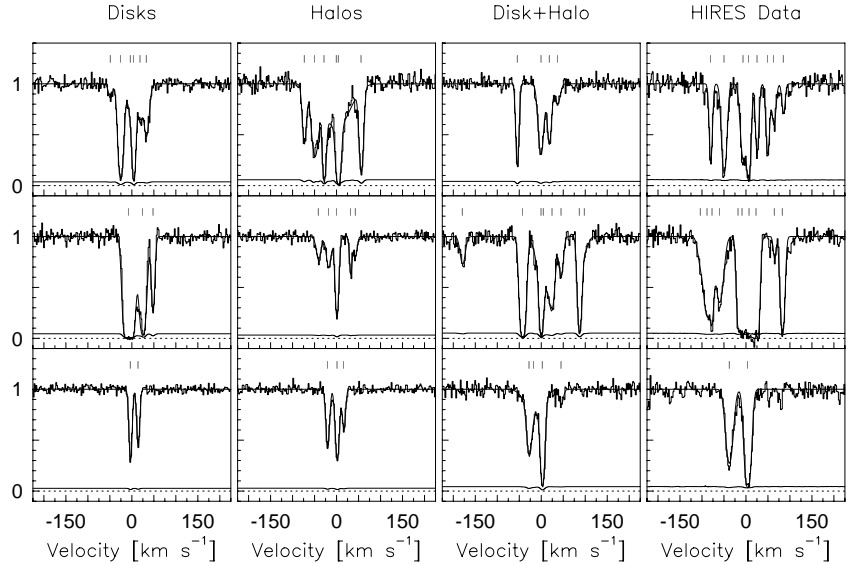}{2.85in}{0}{50.}{50.}{-200}{-23}
\caption{Small sample of model HIRES spectra of the \hbox{{\rm
Mg}\kern 0.1em{\sc ii}} $\lambda 2796$ transition for disk, halo, and
disk+halo models.  The right most panel shows a sample of observed
spectra (see text for discussion).}
\label{cwcfig:profiles}
\vglue -0.2in
\end{figure}

We have observed the \hbox{{\rm Mg}\kern 0.1em{\sc ii}} doublet and
accompanying strong {\rm Fe}\kern 0.1em{\sc ii} transitions with
6.6~km~s$^{-1}$ resolution using the HIRES on Keck (Churchill 1997;
Churchill et~al.\ 1997).
We have observed $\sim 30$ galaxies in absorption at $0.4 \leq z \leq
1.0$.
With HIRES (Vogt 1994), the absorbing gas is seen to resolve into
multiple clouds, each having a well defined velocity, column density
and broadening parameter.
We constructed Monte Carlo model galaxies with disk--only (D),
halo--only (H), and disk+halo (D+H) gas distributions.
D components were given rotational kinematics and H components were
given either infall or isotropic kinematics (cf.~Charlton \& Churchill
1996).
Random ``QSO sight lines'' were then passed through the galaxies and
synthetic HIRES absorption spectra were generated for the \hbox{{\rm
Mg}\kern 0.1em{\sc ii}} doublet and for several strong {\rm Fe}\kern
0.1em{\sc ii} transitions.
These ``spectra'' were analyzed using Voigt profile fitting in
precisely the same fashion as the observational data.
A small sample of model data and the HIRES data are shown in
Fig.~\ref{cwcfig:profiles} for the \hbox{{\rm Mg}\kern 0.1em{\sc ii}}
$\lambda 2796$ transition. 
Ticks mark each cloud's velocity.

\section{Modeling Galaxy Gas Kinematics}

As seen in Fig.~\ref{cwcfig:profiles}, the number of clouds and the
cloud velocities and column densities vary from galaxy to galaxy.
However, in the majority of the galaxies, the gas velocity
distribution suggests a single, dominant absorbing ``complex'' around
which smaller clouds are clustered.
``High velocity" ($v\geq 40$~km~s$^{-1}$) clouds are characterized
by equivalent widths less than 0.1~{\AA} and velocity widths less than
10~km~s$^{-1}$; they are relatively small or low density.
In the final analysis (Charlton \& Churchill 1997), the models that
provide the best statistical match to the data include both the D and H
components, either as a single population of D+H absorbers or a
combined population of D--only and H--only absorbers.
{\it Some \hbox{{\rm Mg}\kern 0.1em{\sc ii}} absorbing gas is arising
in spiral disks.}

\begin{figure}[th]
\cwcpltmacro{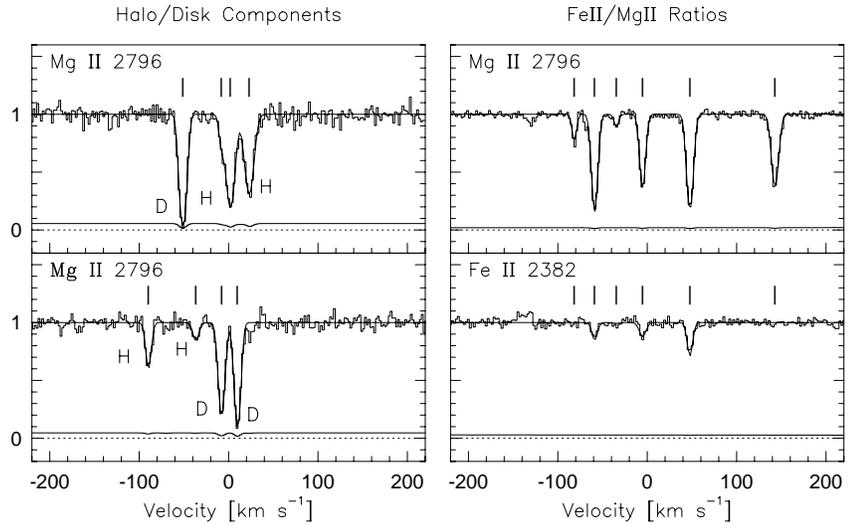}{2.85in}{0}{50.}{50.}{-200}{-23}
\caption{(left) Models of HIRES spectra with disk (D) and halo
(H) clouds labeled.  Zero velocity is given by the optical depth
mean. (right) Observed HIRES \hbox{{\rm Mg}\kern 0.1em{\sc ii}} and
{\rm Fe}\kern 0.1em{\sc ii} profiles at $z=1.32$ showing the
variation in the ratio on a cloud by cloud basis (see text for
discussion).}
\label{cwcfig:festuff}
\end{figure}

\section{Breaking the Disk/Halo Degeneracy}

The difficulty of identifying which absorption lines arise in D or H
components of galaxies is illustrated in the left hand panels of
Fig.~\ref{cwcfig:festuff}.
Is there a technique that can be exploited so that one could securely
infer which clouds arise in D or H components?
Fitzpatrick \& Spitzer (1994) have demonstrated that absorption
properties of \hbox{{\rm H}\kern 0.1em{\sc i}} and \hbox{{\rm H}\kern
0.1em{\sc ii}} regions, and of diffuse gas can be discerned by comparing
the abundance ratios and ionization conditions on a cloud--by--cloud
basis in high resolution spectra.
As illustrated in the right hand panels of Fig.~\ref{cwcfig:festuff},
the column density ratios of {\rm Fe}\kern 0.1em{\sc ii} to
\hbox{{\rm Mg}\kern 0.1em{\sc ii}} are seen to vary from cloud to
cloud, sometimes by a factor of 10 (compare the $v \sim 0$ and
$140$~\hbox{km~s$^{-1}$} clouds).
This ratio is sensitive to the chemical enrichment history of the
particular gas cloud, its ionization condition, and its depletion
pattern.
In the Galaxy, these conditions are seen to be more--less unique
depending upon if they are cool--disk, warm--disk, or halo clouds
(Savage \& Sembach 1996). 
It appears that  with the Galactic ``Rosetta stone'', a deeper
understanding of high redshift galactic gas is forthcoming.

\acknowledgments
CWC thanks the LOC for partial financial support to attend this
Workshop.  This work is supported in part by NSF grant AST--9617185.

\end{document}